# Efficiency of Terrestrial Laser Scanning in Survey Works: Assessment, Modelling, and Monitoring

**Fayez Tarsha Kurdi[1,2]\*, Paul Reed[1], Zahra Gharineiat[2] and Mohammad Awrangjeb[3]**

[1]*East Coast Surveys (Aust) Pty Ltd, Australia*

[2]*Department of Health, Engineering and Sciences, School of Surveying and Built Environment, University of Southern Queensland, Springfield Campus, Australia*

[3]*School of Information and Communication Technology, Griffith University, Australia*

**Submission:** May 03, 2023; **Published:** May 16, 2023

**\*Corresponding author:** Fayez Tarsha Kurdi, School of Surveying and Built Environment, Faculty of Health, Engineering and Sciences, University of Southern Queensland, Springfield Campus, Springfield, QLD 4300, Australia, Email: Fayez.tarshakurdi@usq.edu.au

#### Abstract

Nowadays, static, mobile, terrestrial, and airborne laser scanning technologies have become familiar data sources for engineering work, especially in the area of land surveying. The diversity of Light Detection and Ranging (LiDAR) data applications thanks to the accuracy and the high point density in addition to the 3D data processing high speed allow laser scanning to occupy an advanced position among other spatial data acquisition technologies. Moreover, the unmanned aerial vehicle drives the airborne scanning progress by solving the flying complexity issues. However, before the employment of the laser scanning technique, it is unavoidable to assess the accuracy of the scanner being used under different circumstances. The key to success is determined by the correct selection of suitable scanning tools for the project. In this paper, the terrestrial LiDAR data is tested and used for several laser scanning projects having diverse goals and typology, e.g., road deformation monitoring, building façade modelling, road modelling, and stockpile modelling and volume measuring. The accuracy of direct measurement on the LiDAR point cloud is estimated as 4mm which may open the door widely for LiDAR data to play an essential role in survey work applications.

## Introduction and Related Work

The availability of diverse tools to acquire Light Detection And Ranging (LiDAR) point clouds such as static, mobile, terrestrial, and airborne offer the user to select a suitable option for his project. Despite the difference between the point cloud characteristics such as point density and accuracy according to the employed scanning tools, scanning systems provide almost the same output which is a 3D point cloud, Red, Green, and Blue (RGB), laser intensity and waveform [1]. In fact, the selection of a suitable laser scanning tool depends on the project goal and scale, e.g., to scan a building façade, terrestrial scanning will be better than airborne one because the visibility of vertical elements in airborne scanning will be limited while compared to the terrestrial scanning. Moreover, if the project focuses only on one building façade, the static scanner will be selected, oppositely, if the project goal is to scan all city building facades, in this case, the mobile terrestrial scanner will be the best choice. However, the particularity of each project requires using one or more certain LiDAR data acquisition tool(s).

At this stage, it is important to notice that laser scanning covers a long list of applications such as 3D building modelling, calculation of a Digital Terrain Model (DTM), railway and road monitoring and modelling, powerline modelling, vegetation modelling and biomass estimation, tunnel assessment and orientation, slope and bridge monitoring and pipeline modelling. In fact, the modelling technique differs regarding the target data class, which is why it is unavoidable to classify the LiDAR point cloud before beginning the modelling process. The point cloud classification can be realised by giving each point a label that describes the object that the point belongs to.

Concerning building façade modelling, according to Klimkowska et al. [2], the main utilised raw data to construct building facades are terrestrial LiDAR, images, and both LiDAR and images together. Arachchige et al. [3]; Boulaassal et al. [4], and Previtali et al. [5] derive detailed 3D vector models of building façades starting from terrestrial LiDAR data. They propose applying RANdom SAmple Consensus (RANSAC) algorithm (please see Tarsha Kurdi et al. [6]) to segment the façade into planer patches. Then, starting from the identified planar clusters, façade breaklines are automatically extracted. O'Donnell et al. [7] develop an approach for building façade modelling depending on the angle criterion in boundary detection and the voxelisation representation.

Zhang et al. [8] fuse the LiDAR point cloud with the point cloud extracted from terrestrial façade images to improve the details in the constructed façade model. Chen et al. [9] propose a building façade modelling approach from both LiDAR data and photogrammetric point clouds. The confidence property is employed in the definition of the gradient for each point. Hence, the individual point gradient structure tensor is encoded, whose eigenvalues reflect the gradient variations in the local neighbourhood areas. The critical point clouds representing the building façade boundaries are extracted by analysing gradient structure tensor. To fix the building façade models in an urban scene, particularly for the terrestrial mobile LiDAR data, Zhang et al. [10] suggest a point cloud "fuzzy" repair algorithm based on the distribution regularity of building façade elements.

For road LiDAR data processing, De Blasiis et al. [11] identify and quantify the road degradations of a few types of distresses through a suggested rule-based algorithm. For this purpose, for every single point, the road roughness is supposed as the height deviation which allows recognition of the deformed spots. Puente et al. [12] analysed the deformations of motorway underpasses by analysing a terrestrial LiDAR point cloud.

For road LiDAR data processing, De Blasiis et al. [11] identify and quantify the road degradations of a few types of distresses through a suggested rule-based algorithm. For this purpose, for every single point, the road roughness is supposed as the height deviation which allows recognition of the deformed spots. Puente et al. [12] analysed the deformations of motorway underpasses by analysing a terrestrial LiDAR point cloud.

Zhao et al. [13] suggest an algorithm for extracting street curbs from mobile LiDAR point clouds by applying consecutively three filters which are intensity, elevation, and slope filters to remove buildings and useless parts of point clouds, the invalid slope noise. Finally, they employ X and Y filtered cloud coordinates to establish establish curbs slope function. Ma et al. [14] extract road points by utilising the deep learning network PointNet++(see Qi et al. [15]), afterwards, the road points are processed based on graph-cut and constrained Triangulation Irregular Networks (TIN), and both the commission and omission errors are decreased.Finally, collinearity and width similarity are suggested to approximate the linking probability of road segments. Fernández-Arango et al. [16] propose an approach for pedestrian space extracting and generating a high-definition 3D model. They start by separating terrain and off-terrain classes, and then a K-distance filter is applied to improve and detect the pedestrian spaces.

To measure changes that occur over time, Mahmoud et al. [17] utilise the Digital Terrain Model (DTMs) calculated from airborne LiDAR point cloud for monitoring the Formby sand dunes between 1999 and 2020. Both raster and vector analysis are combined to estimate the dune stability within the focused duration. Miklin et al. [18] study the slope stability by comparing models calculated from LiDAR data and field mapping with available orthophotos of the landslide, resident testimonies, precipitation data, and media releases. Xu et al. [19] use a point cloud measured by Unmanned Aerial Vehicle (UAV) to extract rock parameters and monitor slope stability. Then, the Hough transform is considered to estimate normals for the hue, saturation, and value (HSV) rendering of unstructured point clouds.

To assess the point cloud accuracy, Kersten & Lindstaedt [20] plant a group of control points in the laboratory to assess the scanner accuracy by scanning them and measuring their coordinate directly from the point cloud. Kim et al. [21] perform an accuracy assessment on the LiDAR point cloud to test the presence of any systematic errors. In this context, the absolute vertical accuracy of vegetated and non-vegetated areas is examined. Moreover, both horizontal and vertical absolute errors are also assessed by comparing conjugate points detected from geometric features, e.g., a three-plane feature makes a single unique intersection point which can be computed from the LiDAR point cloud.

This paper represents four different laser scanning projects which are: building façade modelling, road deformation assessment, road modelling, and stockpile scanning and volume calculation. The employed static scanner accuracy is tested. Before starting the modelling step, the pre-modelling step is necessary to extract and filter the useful part of the measured point cloud where the point cloud classification is considered as a part of the pre-modelling step. However, after introducing the paper topic through the introduction section, the pre-modelling section details all requested procedures that come before the modelling step. Thereafter, the modelling section consists of five subsections where the characteristics of the datasets are described and then each project is highlighted and discussed. Finally, the conclusion section draws up the general budget and discusses the envisaged future work.

Before exhibiting the modelling applications, it is important to present the characteristics of the employed datasets.

## Datasets

The terrestrial static Z+F IMAGER 5016 3D laser scanner is utilised to carry out the scans in the projects mentioned in this paper. The main technical characteristics of this system are shown in Table 1. .

At this stage, all acquitted point clouds (see Figure 1a, 2a, 3 & 5) in the next subsections are measured using the scanner illustrated in this subsection. Concerning the point density, theoretically, the point density is variable regarding the overlap or non-overlap area location, the distance from the scanning station and the geometric form of the scanned object. In any way, the mean distance separating points by considering the laser spot diameter (Table 1) is equal to 4mm.

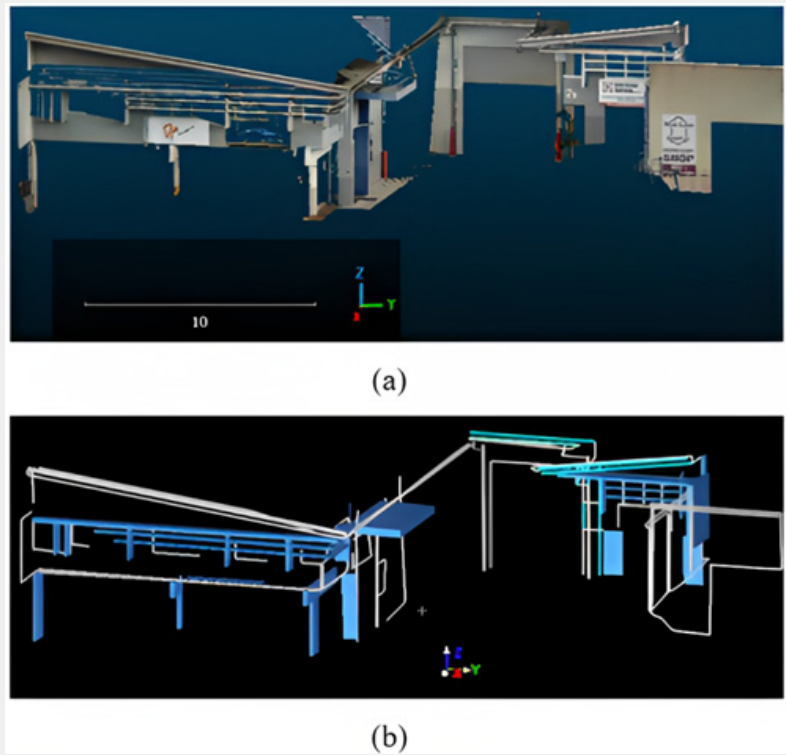

**Figure 1:** (a) 3D façade point cloud; (b) 3D façade model.

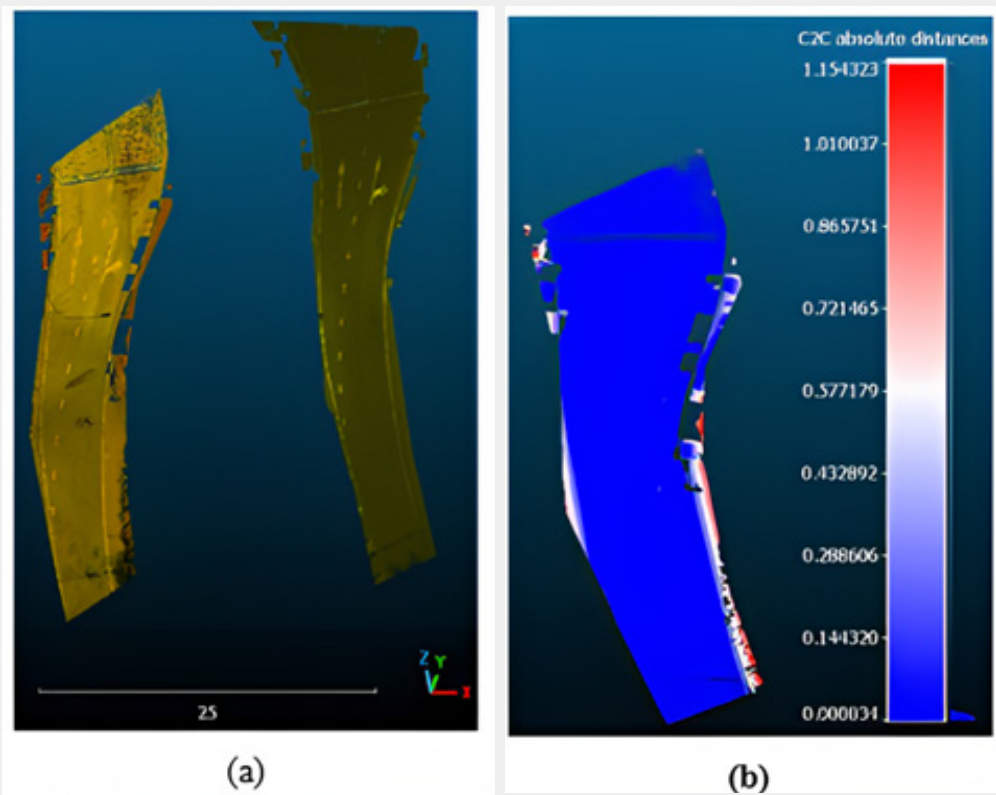

**Figure 2:** (a) Two point clouds of same road (two scans with two weeks shift); (b) Superimposition of two clouds, colours represent vertical distances between them.

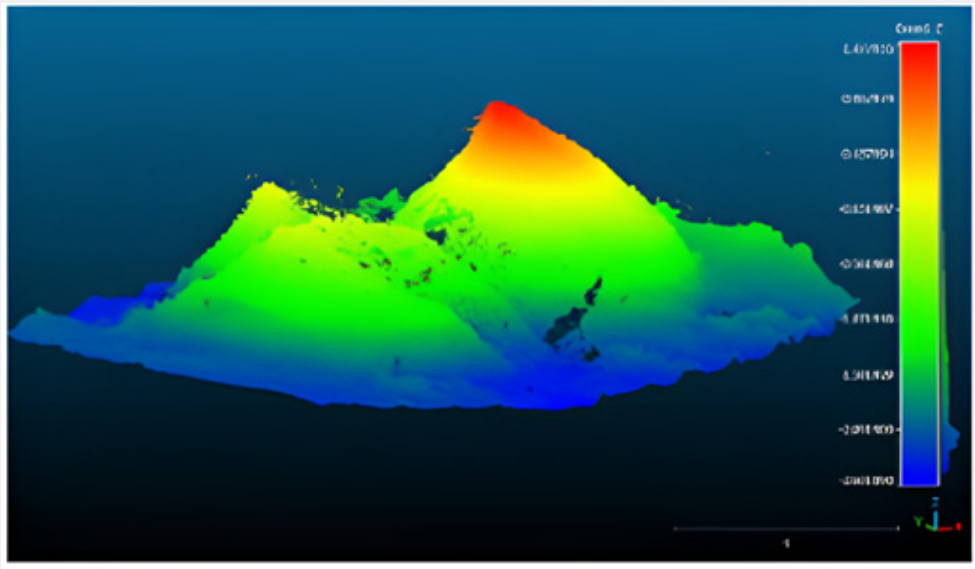

**Figure 3:** 3D visualisation of stockpile DSM.

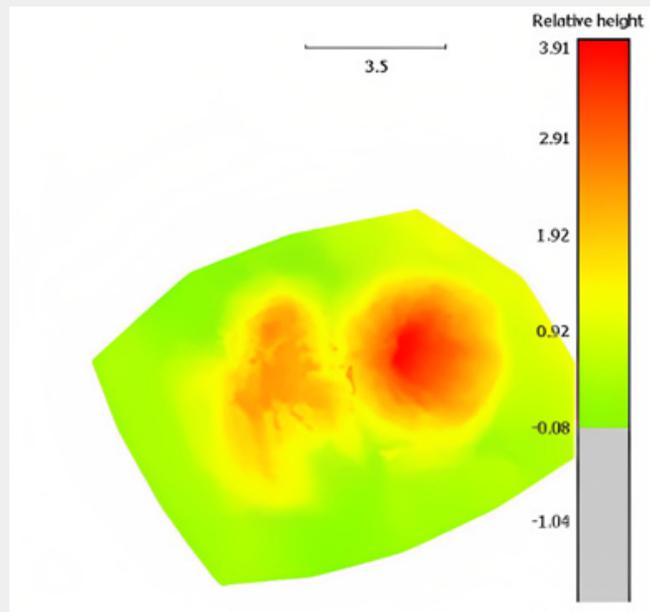

**Figure 4:** 2D visualisation of stockpile DSM considering lowest point height is equal to zero.

**Table 1:** Main Technical Characteristics of Z+F IMAGER 5016 3D Laser Scanner.

| Field-of-view | 360° × 320° |
|---|---|
| Max measurement rate | 1 Mio. points/sec |
| Max range | 360m |
| Laser class | 1 "eye-safe" |
| Operation temperature | -10 °C to +45 °C |
| HDR camera | Full panorama (80 MPixel) |
| Spot diameter | ~ 3.5mm @ 1m / ~ 0.3mrad |

### Pre-Modelling

Once the scan of the target scene is carried out and the requested point clouds are available, it is unavoidable to pre-process them before starting the modelling stage. In fact, according to the project particularity, the requested items from the pre-processing procedure list can be selected and applied. The main procedures of pre-processing step are registration, georeferencing, classification, and filtering. Point cloud registration represents the duty of realising a rigid transformation to accurately align a pair of point clouds [22]. When several point clouds are measured from different viewpoints with overlapping areas, if these point clouds are not georeferenced, it is necessary in this case to transform their coordinate systems to the same one. This operation allows fusing these clouds into one cloud.

Georeferencing is the transformation of the measurements from the sensor coordinate system into an earth-fixed coordinate system [23]. Georeferencing could be achieved during scanning or in the pre-modelling step.

Concerning the point cloud classification, it could be defined as an operation that gives each point a label describing the object that the point belongs to [24]. If the modelling step focuses only on one class such as a building class, in this case, the classification could produce only two classes: buildings and non-building classes. The non-building classes represent all points which do not belong to the building class regardless of their classes. Finally, the filtering operation aims to cancel noisy points which are undesirables during the modelling stage [25]. It is possible to exist a considerable percentage of duplicated points which are due to the overlap areas or the scanning system itself. Hence, the elimination of these points allows for improving the point cloud quality before starting the modelling step.

### Data Modelling

In this paper, four examples of LiDAR data modelling will be presented and discussed: building façade modelling, road deformation monitoring, stockpile volume calculation, and road surface modelling. Moreover, the accuracy of direct measurements on LiDAR point cloud will be assessed. The next subsections exhibit each idea aside.

### Building façade modelling

Building façade point clouds could be considered a rich 3D model because the dense points covering the facades in addition to their RGB, and laser intensity values permit highlighting most of the façade details of different scales. However, extraction of some features from the façade point cloud allows generating a new model that has low memory volume and is easy to be managed. In this context, three feature types are selected to be modelled from the façade point cloud illustrated in Figure 1a. For this peruse, Cyclone 9. 4. 2 software from Leica in addition to Cloudcompare v 2.12 software are used.

The first category is the linear elements such as slot borders (window and door boundaries), plane intersections, and outer boundaries. The second element category is the cylindrical elements such as sewage or drainage pipelines. Finally, solid elements such as columns, umbrellas, and decoration parts are also considered (Figure 1b). To summarise, before constructing a façade model from the LiDAR point cloud, the requested façade elements should be determined. Though the constructed model will be lighter regarding the memory volume viewpoint, the original point cloud is still representing the richest detailed façade model.

### Road deformation monitoring

A sewage line has been placed under the roadway through under bore, and part of the conditions required monitoring of the road surface for any deformation. For this purpose, two scans are performed for the road section with two weeks shift between them. To calculate the vertical distances between the two obtained point clouds, the two point clouds should be registered after filtering them, especially by eliminating the moving vehicle points (Figure 2a). Therefore, the road marks are used to achieve the registration (cloud to cloud registration). Figure 2b shows the visualisation of vertical distances distances the two road point clouds. It can be noticed that road deformations are within the accepted limits (~ 0m). Indeed, all road body is illustrated in blue colour which means according to the colour scale shown on the right side of Figure 2b that the deformations are around zero value. The same work will be repeated several times with different time shifts to ensure road safety.

At this stage, it is important to refer to the possibility of applying the same test to monitor one slope's stability. The difference between the two cases is that in the case of the slope, it is expected that the terrain will be covered by vegetation which must be eliminated in the filtering step [26,27]. Finally, all monitoring projects are based on the same idea which is achieving several scans of the interest object with time shifts and then carrying out a positional comparison between the measured LiDAR point clouds.

### Stockpile volume

In a given civil engineering project, it was requested to measure the volume of a topsoil stockpile. In this project, the stockpile is scanned from four different viewpoints. Spherical targets were placed in fixed locations which were read from several of the scan set ups. The registration step is carried out during the scanning using these spherical targets. The stockpile point cloud is extracted and then filtered from the whole measured point cloud. Figure 3. mentions the 3D visualisation of the stockpile Digital Surface Model (DSM). In this model, the missing points which represent hidden points those could be calculated by interpolating the neighbouring points [28] (Figure 4). To calculate the stockpile volume, it is assumed that the lowest point height in the scanned scene is equal to zero (Figure 4), then the pixel values are recalculated according to this hypothesis. Afterwards, each pixel represents a cuboid, its base area is equal to the pixel area (the resolution), and its height is equal to the pixel value. The summation of all cuboid volumes provides the stockpile volume. In this way, the user can modify the lowest pixel value according to the project goal. In the example presented in Figure 3 & 4, the stockpile volume is equal to 140.86m3 and the area is equal to139.11m$^2$.

### Road surface modelling

In road survey work, not only road deformation monitoring could be achieved using laser scanning technology, but road modelling and feature extraction could be also done. In a road survey, the point cloud georeferencing is unavoidable, that is why three known coordinate targets (control points) in minimum should be used in the scanning project. Furthermore, according to the number of the installed scanning stations, additional targets must be used to create links between consecutive stations to register the measured point clouds. In the project presented in Figure 5, the Triangulated Irregular Network (TIN) is calculated from the filtered road point cloud, and then superimposed over the LiDAR point cloud. However, regarding the high point cloud accuracy (less than 10mm, please see Table 1), other operations could be carried out on the road point cloud such as curb-line extraction, tree, signs, and light locations.

At this stage, it is important to refer that the two main disadvantages of terrestrial laser scanning in road work are the short efficient distance between two consecutive scanning stations (35m) and the presence of hidden areas in the point cloud. To overcome the last two issues, the number of scanning stations must be increased. In the project presented in Figure 5, only one scanning station is installed in the open and small area of the project zone.

### Scanning accuracy estimation

In laser scanning, three types of accuracies could be distinguished: the accuracy of point coordinate calculation, which is provided by the scanner company, and the accuracy of direct measurement in the point cloud, which can be estimated on the site or in the laboratory [20], and finally the accuracy of the calculated model from the point cloud [24]. This paper focuses on the accuracy of direct measurements on the point cloud, that is mean any direct measurement achieved on the point cloud to extract some information such as distance, area, coordinates, and volume. To test the accuracy of the direct measurements on the point cloud, there is no-standard method to achieve this goal [20], that is why the next experiment is suggested and realised as follows.

Four targets are installed in Points 1, 2, 3, and 4 according to Figure 6. Thereafter, the scanner is installed on Point 0. Four scans are realised for the targets listed above. Then, the distances between targets are measured directly from the four obtained point clouds. Table 2 represents the average distance values between the targets among the four scans. Table 3 shows the standard deviation values of the last measured distances.

**Table 2:** Average distance values between scanner and targets through four scans.

| Point Number | 1 | 2 | 3 | 4 |
|---|---|---|---|---|
| 1 | | 7.235 | 12.863 | 18.325 |
| 2 | | | 6.385 | 13.436 |
| 3 | | | | 7.715 |
| 4 | | | | |
| | Average distances (m) | | | |

**Table 3:** Standard Deviation Values of Average Distances Represented in Table 2.

| Point Number | 1 | 2 | 3 | 4 |
|---|---|---|---|---|
| 1 | | 2.4 | 2.8 | 0.8 |
| 2 | | | 1.7 | 3.4 |
| 3 | | | | 3.9 |
| 4 | | | | |
| | Distance Standard Deviation (mm) | | | |

It can be noticed from Table 3 that the maximum obtained value of standard deviations is equal to 3.9mm. This result reflects the high accuracy of the employed scanner. Furthermore, this accuracy makes the scanning qualified to be integrated within the survey work.

### Conclusion

This paper has presented four projects based on the use of terrestrial LiDAR point clouds. First, building facade modelling was realised by assuming that one façade consists of three main geometric elements which are straight lines, cylinders, and solid elements. Second, road deformation monitoring was carried out by superimposing two point clouds of the same road which were measured with time spacing, and then the distances between the two point clouds were measured. Third, the hips of depresses were scanned and DSM was calculated to measure the volume of the stockpile. Finally, in the context of road modelling, one road was scanned, thereafter the obtained point cloud was filtered, and the TIN structure of the concerned road was calculated. It was unavoidable to test the scanner's accuracy to validate the realised work. In fact, the high accuracy of direct measurement from the acquired point clouds (average value = 4mm), in addition to the available automatic and semi-automatic LiDAR data processing tools give the priority to laser scanning technology. The listed applications in this paper confirm the successful selection of the terrestrial scanner as a tool for data acquisition. Moreover, the LiDAR data efficiency was proved for achieving the project targets. Finally, in future work, it is confronted to realise most of the expected applications using different sources of LiDAR data simultaneously or in an individual way. The more efficient and available LiDAR data processing tools will be tested and listed.

## Acknowledgment

All datasets were measured in Queensland, Australia were provided by East Coast Surveys (Aust) Pty Ltd, http://www.eastcoastsurveys.com.au.

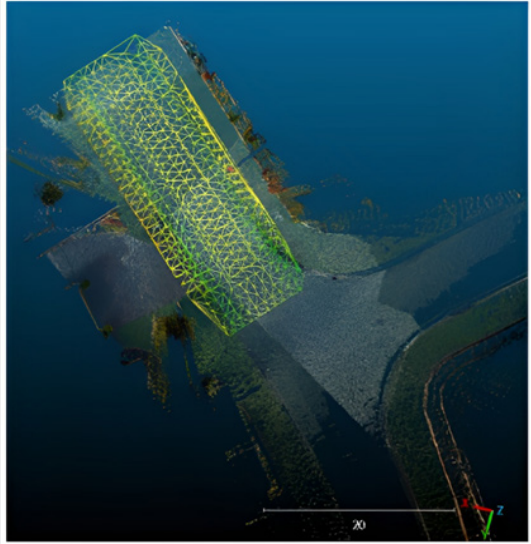

**Figure 5:** Superimposition of calculated TIN on road point cloud.

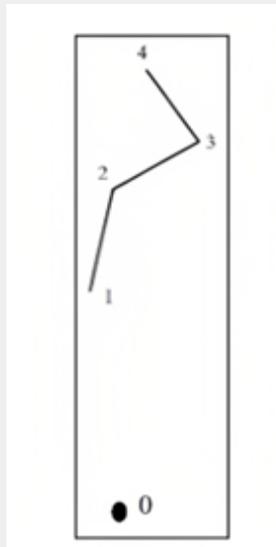

**Figure 6:** Point (0) is scanner location, Points (1, 2, 3, and 4) are locations of targets.